\documentclass{jpsj2}
\usepackage{graphicx}

\newcommand{\bq}{\mbox{\boldmath $q \:$}}
\newcommand{\bI}{\mbox{\boldmath $I \:$}}

\newcommand{\bH}{\mbox{\boldmath $H \:$}}
\newcommand{\brho}{\mbox{\boldmath $\rho \:$}}

\newcommand{\BR}{\mbox{\boldmath $R \:$}}
\newcommand{\Bq}{\mbox{\boldmath $q \:$}}
\newcommand{\BQ}{\mbox{\boldmath $Q \:$}}
\newcommand{\BJ}{\mbox{\boldmath $J \:$}}
\newcommand{\BI}{\mbox{\boldmath $I \:$}}

\def\Beqa{\begin{eqnarray}}
\def\Eeqa{\end{eqnarray}}
\def\H{\hspace{0.5cm}}

\def\EX{{\rm e}}
\def\I{{\rm i}}
\def\TH{\theta}

\title{%
Invariant Form of Hyperfine Interaction with Multipolar Moments\\
 - Observation of Octupolar Moments in  NpO$_{2}$ and  CeB$_{6}$ by NMR -
}

\author{%
Osamu \textsc{Sakai}\thanks{E-mail: sakai@phys.metro-u.ac.jp.}, 
Ryousuke \textsc{Shiina}
and Hiroyuki \textsc{Shiba}$^1$
}

\inst{%
Department of Physics, 
Tokyo Metropolitan University, Hachioji 192-0397, Japan \\
$^{1}$The Institute of Pure and Applied Physics, 6-9-6 Shinbashi, Tokyo 105-0004, Japan \\
}

\recdate{
}

\abst{%
The invariant form of the hyperfine interaction between multipolar 
moments  and  
the nuclear spin 
is derived,  and  applied to discuss possibilities to identify the antiferro-octupolar (AFO) moments
by  NMR experiments. The  ordered phase of NpO$_{2}$ and the phase IV of Ce$_{1-x}$La$_{x}$B$_{6}$ are studied in detail. 
Recent  $^{17}$O NMR for polycrystalline samples of 
NpO$_{2}$ are discussed theoretically from our formulation. 
The observed feature of the splitting of $^{17}$O NMR spectrum into a 
sharp  line and a broad line,  their intensity ratio, and 
the magnetic field dependence of the shift and of the width can be 
consistently explained on the basis of the triple $\bq$ AFO ordering 
model proposed by Paix\~{a}o {\it et. al.} Thus, the present theory shows that  the $^{17}$O NMR spectrum gives a strong support to the model. 
The 4 O sites in the fcc NpO$_2$ become inequivalent due to  the secondary triple $\bq$ ordering of AF-quadrupoles:  one cubic and three non-cubic sites. It turns out that the hyperfine field due to the antiferro-dipole and AFO moments induced by the magnetic field, and the quadrupolar field at non-cubic sites are key ingredients to understand the observed spectrum.
The controversial problem of the nature of phase IV in Ce$_{1-x}$La$_{x}$B$_{6}$ is also studied. It is pointed out that there is a unique feature in the NMR spectra, if the 
$\Gamma_{5}$($T^{\beta}_{x}=T^{\beta}_{y}=T^{\beta}_{z}$) AFO ordering 
is realized in Ce$_{1-x}$La$_{x}$B$_{6}$. Namely,
the hyperfine splitting of a B atom pair on the 
$(\frac{1}{2},\frac{1}{2},\pm u)$ sites crosses zero 
on the $(1\bar{1}0)$
plane when the magnetic field is rotated around
the $[001]$ axis. 
}

\kword{
antiferro-octupolar order, octupolar moment,
hyperfine interaction, NMR,  NpO$_{2}$,  CeB$_{6}$,
antiferro-quadrupolar order
}

\begin{document}
\sloppy
\maketitle

\section{Introduction}

The antiferro-octupolar (AFO) moments, which are odd in time-reversal symmetry but are distinct  from dipole moments,  are now widely recognized as a ``hidden'' order parameter\cite{A1}. The AFO order is a promising candidate to resolve the breaking of the time reversal symmetry without the antiferromagnetic dipole (AFM) order. However, a direct observations of the AFO is very difficult. 

There are two ways to realize the AFO moments. One is induced AFO moments in the 
antiferro-quadrupolar (AFQ) state by applying the static 
magnetic field\cite{A2}. Another is a spontaneous AFO ordering, which has been recently proposed
for NpO$_2$ and Ce$_{1-x}$La$_{x}$B$_{6}$\cite{B0,A5,A6,A7,A3,A4}. 
In this paper, we wish to study a possibility to detect the AFO moments in NMR experiments through the hyperfine interaction. 
For this purpose we derive a general form of the hyperfine interaction
between the multipolar moments and the nuclear spin. 
Actually we studied this problem before for CeB$_6$. The approach was to derive a phenomenological form of the hyperfine interaction from the symmetry \cite{A8,A9}. 
However, there is a simpler and equivalent approach, which we use in this work.  It is a straightforward generalization of  ref.\citen{A10} , in which  the invariant coupling form between 
multipolar moments was derived by using a method of symmetrized molecular orbital theory.
The usefulness of this approach to discuss the hyperfine interaction will be demonstrated in this paper. 

To have octupolar moments as independent local degrees of freedom, the material must have high symmetry such as cubic. There are three cubic systems for which the AFO order of some sort is either realized or is likely to be realized.  The first system is CeB$_{6}$, which has been studied extensively\cite{A9,B1,B2,B3}.  There are a number of anomalous features in the phase diagram as well as in  the nature of  the low-temperature order phases, phase II in particular.  They are naturally explained by noting  the
interaction effect between the {\it field-induced AFO moments}, especially of the
$T_{xyz}$ type ($\Gamma_{2}$ type)\cite{A9,A12}.  However, a  crucial progress was achieved, when  the 
apparent  inconsistency between the neutron diffraction\cite{B6}
and NMR\cite{B7} for the phase II was successfully resolved by recognizing that  the hyperfine 
coupling is present between  the B nuclear spin and  the induced $T_{xyz}$ AFO moment of Ce ion\cite{A9,A8,A13}.  This is the first case in which the contribution from AFO moments  to the hyperfine coupling has been discovered.  Therefore the NMR is now regarded as a direct proof for the field-induced AFO moment in CeB$_6$.

The second system is NpO$_2$; the nature of the order in NpO$_2$ below
26K was a mystery for a long time. Carrying out a resonant X-ray
scattering,  Paix\~{a}o {\it et. al.} have recently suggested that their
results can be explained well by assuming the triple $\bq$ AFQ order of
$\Gamma_{5}$ type\cite{A4}.
They proposed the triple $\bq$ AFO as the primary
order parameter because any dipole order have not been observed 
though  Np$^{4+}$ is the  Kramers ion. The triple $\bq$ AFQ is induced from the primary triple $\bq$ AFO.  A new development on this problem has been brought by  
Tokunaga {\it et. al.}'s $^{17}$O  ($I=5/2$) NMR experiment 
for polycrystalline samples\cite{A15}. 
They found that the O ion sites, which are equivalent in the 
high temperature phase ($Fm\bar{3}m$), split into two 
inequivalent groups. The experimental results have the following unique features: 
\parindent 0pt
\par
(I) 1/4 of the$^{17}$O nuclei  contributes to a sharp resonance line, while the rest 3/4 gives a broad resonance line.

(II) The shift of the sharp line is approximately proportional to the strength of the applied magnetic field. 

(III) The width of the broad line is given as the sum of a constant and a part proportional to the strength of the magnetic field.

(IV) The shift of the sharp line and the magnetic field induced width of the broad line have comparable magnitude. 
\par
\parindent 10pt

We show in this paper that those experimental results can be naturally explained by the 
quadrupolar field and the  
hyperfine field splitting  in the triple $\bq$ state. 
The hyperfine field is caused by the 
AFO and/or AFM moment induced by the application of the magnetic 
field. Therefore the $^{17}$O NMR is a fingerprint of the triple $\bq$ order of AFO and AFQ. 

The third system is Ce$_{x}$La$_{1-x}$B$_{6}$. It is known that  different from pure CeB$_6$ a new phase, which is called phase IV, appears instead of phase II in the low magnetic field region of
Ce$_{x}$La$_{1-x}$B$_{6}$ ($x \sim 0.75$)\cite{C1,C2,C3}. The nature of phase IV remains controversial.  The time reversal symmetry breaking 
is reported for this phase\cite{A11,B4}. 
However, no indication of the AFM order has been 
detected experimentally, although  neutron diffraction studies were carried out extensively\cite{B5}.
A characteristic cusp of the magnetic susceptibility has been observed
at the transition temperature between phase I and phase IV\cite{A7}.
From  these results together with  other experimental indications\cite{C1,C2,C3,D1,D2}, an 
AFO order of the $\Gamma_{5}$ type has been recently 
proposed  for the phase IV\cite{A5,A6,A7}.  A recent NMR experiment indicated a broadening of NMR lines in the phase IV, showing a sign of time reversal symmetry breaking\cite{A11}. No further information is obtained from this NMR experiment. 
Quite recently,  the importance of the uniaxial stress 
to remove the ambiguities due to the domain structure has been demonstrated. 
Morie {\it et. al.} pointed out that the magnetic susceptibility shows an 
anisotropy in contrast to the isotropic behaviors reported previously for ambient 
pressure\cite{A14}. 
This result is qualitatively similar to what one obtains theoretically from  the AFO of the $\Gamma_{5}$ type with 
$\langle T^{\beta}_{x}\rangle =\langle T^{\beta}_{y}\rangle=\langle T^{\beta}_{z}\rangle$.
Here, $T^{\beta}_{x}$, $T^{\beta}_{y}$ and $T^{\beta}_{z}$ are the three components of the $\Gamma_{5}$ type octupolar moment\cite{G1}. 
The NMR study under uniaxial stress is desirable to clarify the situation for phase IV.
In this paper we will point out a  possibility to identify the $T^{\beta}$ type AFO order
through the NMR experiment.  We will show that if  the $\langle T^{\beta}_{x}\rangle=\langle T^{\beta}_{y}\rangle=\langle T^{\beta}_{z}\rangle$ 
type order is realized as Kubo and Kuramoto, and Morie {\it et. al.} 
assumed\cite{A5,A14},  a characteristic feature of the NMR splitting should be present 
in the field-direction dependence of the B NMR from  ($\frac{1}{2},\frac{1}{2},\pm u$) sites. 

This paper is organized as follows. 
In \S 2, the hyperfine interaction with multipolar moments of 4f
electrons is derived for NpO$_{2}$
on the basis of new method.
In \S 3,   the
recent $^{17}$O NMR experiment on NpO$_{2}$ is studied theoretically 
to show that it is consistent with the triple $\bq$ structure of AFO and AFQ. 
In \S 4, the hyperfine interaction 
in CeB$_{6}$ is re-derived.
The
field direction dependence of the NMR line is discussed in \S 5, 
having the phase IV problem in our mind. 

\section{ Invariant Coupling in NpO$_{2}$ and Splitting of O sites 
into  Non-equivalent Sites
}

The Np ions in NpO$_{2}$ form the f.c.c. lattice.
The position vector  $\brho$ of the eight O ions in the f.c.c. cube is given in Table \ref{t1},
where the length of the edge of the cube is chosen as $2a$.
The local symmetry around the O sites has the T$_{d}$ character.
From the discussions described in Appendix A, the  hyperfine interaction of the $^{17}$O nuclear  spin ($I=5/2$)  located  at $\brho$  with multipolar moments of Np 5f electrons  is given by 
\Beqa
H_{\rm hf}(\brho)
 & = &
\EX^{\I\Bq\brho}
I_{x}(\brho) 
\Big[
c_{1,1}
\frac{4}{\sqrt{4}}
J_{x}(\Bq)
\big( c_{x}c_{y}c_{z}-\I sg({\brho})s_{x}s_{y}s_{z} \big)
\nonumber \\
 &  &
\hspace{30pt} +c_{1,2}\frac{4}{\sqrt{16}}
\Big\{  J_{y}(\Bq) \big(\I sg(\brho) s_{z}c_{x}c_{y}-c_{z}s_{x}s_{y} \big)
\nonumber \\
 &  &
\hspace{70pt} + J_{z} (\Bq) \big(\I sg(\brho) s_{y}c_{z}c_{x}-c_{y}s_{z}s_{x} \big) \Big\}
\nonumber \\ 
&  &
\hspace{30pt}  +
 c_{1,3}\frac{4}{\sqrt{8}}
 \Big\{  T^{\beta}_{y} \big(\Bq) \Big\{ \I sg(\brho) s_{z}c_{x}c_{y}-c_{z}s_{x}s_{y}\big)
\nonumber \\
 &  &
 \hspace{70pt}  -
  T^{\beta}_{z}(\Bq)\big (\I sg(\brho) s_{y}c_{z}c_{x}-c_{y}s_{z}s_{x})\big) \Big\}
\nonumber \\
 &   & 
 \hspace{30pt} +
 c_{1,4}\frac{4}{\sqrt{4}}
 T_{xyz}(\Bq) \big(\I sg(\brho)
 s_{x}c_{y}c_{z}-c_{x}s_{y}s_{z} \big) 
\Big] 
\nonumber  \\
& & + ({\rm cyclic}\   {\rm permutation} \  {\rm of} \  x,  y\   {\rm and} \  z)\ . 
\label{eq2.1} 
\Eeqa
Here $c_{i,j}$'s  are coupling constants, 
and 
$
 sg(\brho)=(-1)^{(\rho_{x}+\rho_{y}+\rho_{z}-\frac{a}{2})/a}
$,
and  $c_{\nu}$($s_{\nu}$) represents $\cos( q_{\nu}/2)$
($\sin( q_{\nu}/2)$). 
The summation over $\Bq$ is assumed in eq. (\ref{eq2.1}).
Under the cubic  rotation group 
the octupolar operator of $\Gamma_{4}$ type 
has the same symmetry character with the dipole operator.  Therefore the dipole operator $\BJ$ in eq. (1) should be read as 
a linear combination of the pure dipole and the octupolar 
operator of $\Gamma_{4}$ type.
The terms containing  the factor
$\I sg(\brho)$ 
are due to the 
T$_{d}$ site symmetry around each O ion. 
The imaginary factor in this quantity is only a seeming, which is
compensated by the factor
$\EX^{\I\Bq\brho}$.

Let us assume the $\Gamma_{5}$  AFO order  of
triple $\bq$ type:
$T^{\beta}_{x}(\BR)
=T^{\beta}_{x}\EX^{\I\BQ_{x}\BR}$,
$T^{\beta}_{y}(\BR)
=T^{\beta}_{y}\EX^{\I\BQ_{y}\BR}$
and 
$T^{\beta}_{z}(\BR)
=T^{\beta}_{z}\EX^{\I\BQ_{z}\BR}$,
with
$\BQ_{x}=\pi(1,0,0)$,
$\BQ_{y}=\pi(0,1,0)$ and 
$\BQ_{z}=\pi(0,0,1)$,
as assumed in ref. (\citen{A4} ).
Then we easily find from eq. (\ref{eq2.1}) that the AFO moments in this order do not 
induce any  hyperfine field 
on the O atoms.

As noted in refs. (\citen{A4}) and (\citen{ A10}) 
and seen from Table \ref{t0},
the triple $\bq$ ordering induces the AFQ of the $\Gamma_5$ type: 
$O_{yz}(\BR)
=O_{yz}\EX^{\I\BQ_{x}\BR}$,
$
O_{zx}(\BR)
=O_{zx}\EX^{\I\BQ_{y}\BR}$ and
$
O_{xy}(\BR)
=O_{xy}\EX^{\I\BQ_{z}\BR}$.
If $O_{yz}=O_{zx}=O_{xy}\equiv O_{\Gamma_{5}}$ hold, 
the cubic symmetry of the crystal is not broken.

Let us consider the quadrupolar interaction between the 
$^{17}$O nuclear moment 
on $\brho$ and the
quadrupolar moment 
of Np ion. 
The interaction form is given by following the discussion in 
Appendix~B: 
\Beqa
H_{\rm qq}(\brho)
   = &
\EX^{\I\brho\Bq}
\Big[c_{2,1}\frac{4}{\sqrt{4}}
\big((O_{u}(\brho)O_{u,\Bq}+O_{v}(\brho)O_{v,\Bq})
(c_{x}c_{y}c_{z}-\I sg(\brho)s_{x}s_{y}s_{z})\big)
\nonumber \\
   + &
c_{2,2}\frac{8}{\sqrt{24}}
\big(
\frac{1}{2}(
-O_{u}(\brho)
+\sqrt{3}O_{v}(\brho))
O_{yz,\Bq}
( - \I sg(\brho) s_{x}c_{y}c_{z}+c_{x}s_{y}s_{z})
\nonumber \\
 &
+O_{u}(\brho)
O_{xy,\Bq}
(- \I sg (\brho) s_{z}c_{x}c_{y}+c_{z}s_{x}s_{y})
\nonumber \\
 &
+\frac{1}{2}(
-O_{u}(\brho)
-\sqrt{3}O_{v}(\brho))
O_{zx,\Bq}
(-\I sg(\brho) s_{y}c_{z}c_{x}+c_{y}s_{z}s_{x})\big)
\nonumber \\
 + &
c_{2,3}\frac{4}{\sqrt{4}}
\big(O_{yz}(\brho)\frac{1}{2}(-O_{u,\Bq}+\sqrt{3}O_{v,\Bq})
(\I sg(\brho)
 s_{x}c_{y}c_{z}-c_{x}s_{y}s_{z}
)
\nonumber \\
  &
+ O_{zx}(\brho)\frac{1}{2}(-O_{v,\Bq}-\sqrt{3}O_{v,\Bq})
(\I sg(\brho)
 s_{y}c_{z}c_{x}-c_{y}s_{z}s_{x})
 \nonumber \\
 &
+O_{xy}(\brho)O_{u,\Bq}
(\I sg(\brho)
 s_{z}c_{x}c_{y}-c_{z}s_{x}s_{y}
)\big)
\nonumber \\
 + &
c_{2,4}\frac{4}{\sqrt{4}}
\big((O_{yz}(\brho)O_{yz,\Bq}
 +O_{zx}(\brho)O_{zx,\Bq}
 +O_{xy}(\brho)O_{xy,\Bq})
(c_{x}c_{y}c_{z} + \I sg(\brho)s_{x}s_{y}s_{z})
\big)
\nonumber \\
 + &
c_{2,5}\frac{4}{\sqrt{8}}
\Big(O_{yz}(\brho)\big(
O_{zx,\Bq}(\I sg(\brho)s_{z}c_{x}c_{y}-c_{z}s_{x}s_{y})
+O_{xy,\Bq}(\I sg(\brho)s_{y}c_{z}c_{x}-c_{y}s_{z}s_{x})\big)
\nonumber \\
  &
+({\rm cyclic} \hspace{0.2cm} {\rm permutation} \hspace{0.2cm} 
{\rm of} 
\hspace{0.2cm} x, y \hspace{0.2cm} {\rm and}  
\hspace{0.2cm} z 
)
\Big)
\Big]
.
\label{eq2.2}
\Eeqa
We can derive the quadrupolar interaction between the nuclear quadrupole of
$^{17}$O and the 
triple $\bq$ AFQ of Np 5f electrons as 
\Beqa
H_{\rm qq}(\brho,c_{2,2})
  & = &
C_{2,2}
( - \I sg(\brho)) 
\Big[
\frac{1}{2} \Big(
-O_{u}(\brho)
+\sqrt{3}O_{v}(\brho) \Big)
O_{yz}\EX^{\I\BQ_{x}\brho}
\nonumber \\
 & &
+O_{u}(\brho)
O_{xy}\EX^{\I\BQ_{z}\brho}
+\frac{1}{2} \Big(
-O_{u}(\brho)
-\sqrt{3}O_{v}(\brho) \Big)
O_{zx}\EX^{\I\BQ_{y}\brho}
\Big],
\label{eq2.3}
\Eeqa
where the quantity $C_{2,2}$  is proportional to  $c_{2,2}$ and the 
order parameter of AFQ $O_{\Gamma_{5}}$, or square of the 
order parameter of AFO $( T^{\beta})^{2}$, i.e.
$C_{2,2} \propto c_{2,2}O_{\Gamma_{5}} \propto
 c_{2,2} ( T^{\beta})^{2}$.
This interaction  originates from the coupling caused by
 the T$_{d}$ symmetry, such as $s_{z}c_{x}c_{y}$ term in eq.(\ref{eq2.2}).
At the $\brho_{1}=\frac{1}{2}(3a,3a,3a)$ site,
the factor
$
-\I sg(\brho_{1})
\EX^{\I\BQ_{\nu}\brho_{1}}
$ 
takes 1 for all $\nu$.
Therefore 
it gives the following expression
\Beqa
H_{\rm qq}(\brho_{1},c_{2,2})
  & = &
-C_{2,2}
\Big[
 O_{u}(\brho_{1})
(-\frac{1}{2}O_{yz}+O_{xy}-\frac{1}{2}O_{zx})
\nonumber \\
 & &
\hspace{20pt} +\frac{\sqrt{3}}{2}
O_{v}(\brho_{1})(O_{yz}-O_{zx})
\Big].
\label{eq2.4}
\Eeqa
If  $O_{yz}=O_{zx}=O_{xy}\equiv O_{\Gamma_{5}}$ holds, 
this interaction is canceled out on the $\brho_{1}$ site.

On the other hand,  at the site 
$\brho_{2}=\frac{1}{2}(3a,3a,a)$,
the factor take $-1, -1$ and $1$, respectively, for
the terms characterized by the vectors, $\BQ_{x},\BQ_{y}$ and $ \BQ_{z}$.
Then we have 
\Beqa
H_{\rm qq}(\brho_{2},c_{2,2})
  & = &
C_{2,2}
\Big[
 O_{u}(\brho_{2})
\Big(-\frac{1}{2}O_{yz}-O_{xy}-\frac{1}{2}O_{zx} \Big)
\nonumber \\
  &  &
\hspace{20pt} +
\frac{3}{2}O_{v}(\brho_{2})(O_{yz}-O_{zx})\Big],
\nonumber \\
   & = &
C_{2,2}  O_{u}(\brho_{2})(-2O_{\Gamma_{5}}).
\label{eq2.5}
\Eeqa
Therefore the quadrupolar field remains on this O atom.
A similar calculation can be carried out for other sites and the result is 
shown in Table \ref{t2}.
(The sign of $-\I sg(\brho)\EX^{\BQ_{\nu}\brho}$ is given in the Table
\ref{t1}.)
We note $O_{u}$ of the $^{17}$O nucleus ($I=5/2$) is proportional to 
$3I^{2}_{z}-I(I+1)$, and 
$-\frac{1}{2}(O_{u}+\sqrt{3}O_{v}) \propto 3I^{2}_{y}-I(I+1)$, and 
$-\frac{1}{2}(O_{u}-\sqrt{3}O_{v}) \propto 3I^{2}_{x}-I(I+1)$.
The triple $\bq$ AFQ order causes 4 different quadrupolar fields 
on the eight O sites in the fcc cube: a pair of  cubic sites 
(with zero quadrupolar field) and 3 pairs of  uniaxial-symmetry 
sites (
each of which has principal axis along one of the x,y and z axes).
The $^{17}$O nucleus on the non-cubic site will show the 
the quadrupolar splitting of the spectrum.
The appearance of the cubic and non-cubic site in the 
triple $\bq$ AFO state can be understood in an intuitive way 
as explained in Figs. 1 and 2.
The colored cubes are the cubic sites and their neighboring 
three sites have principal axis along the x, y  and z axes.

The quadrupolar field of the type $O_{yz}$, $O_{zx}$ and $O_{xy}$
on O atoms is not induced in the
triple $\bq$  ordering.

\begin{figure}[htb]
\begin{center}
\includegraphics[width=8cm]{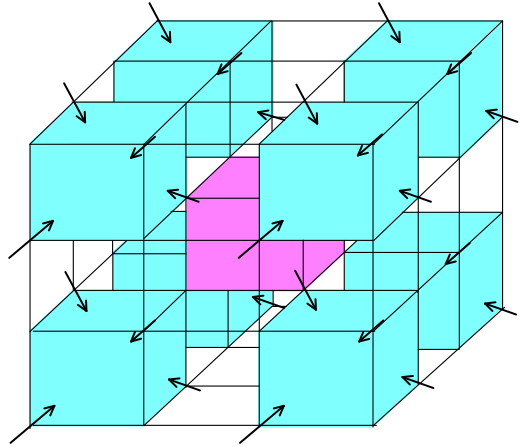}
\end{center}
\caption{Schematics for the triple $\Bq$ ordering of $T^\beta$ AFO. The
$T^\beta$
moments on
Np ions are denoted by arrows.
The O sites are located at the centers of colored and colorless cubes.
The unit cell of the ordered phase is composed of four Np and eight O
sites.
See also Fig. 2 for details.
}
\label{F1}
\end{figure}

\begin{figure}[htb]
\begin{center}
\includegraphics[width=8cm]{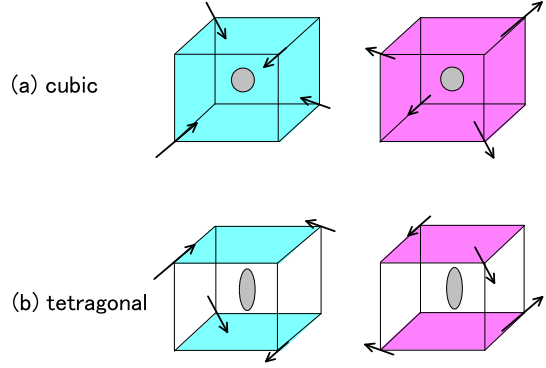}
\end{center}
\caption{
Environments around O sites located at the center of cubes.
There are eight nonequivalent O sites.
(a) Two of them are subject to cubic fields. Four arrows point the
center
of the blue cube, whereas they go out from the center of the red cube.
Both
keep
four trigonal axes, so that a cubic symmetry is retained there.
(b) Two O sites in the colorless cubes are neighboring with blue and
red
colored cubes
along z axis. Thus, the trigonal symmetry is broken there, and they
are
subject to
tetragonal fields. There are other four O sites neighboring with
colored
cubes
along $x$ and $y$ axes, respectively. These can be represented by the
corresponding
rotations of the figure (b).
}
\label{F1}
\end{figure}

\begin{table}[t]
\caption{Product of $\Gamma_{5}$ octupolar moment arranged as the 
quadrupolar moments.
}
\label{t0}
\begin{halftabular}{@{\hspace{\tabcolsep}\extracolsep{\fill}}cccc}
\hline
$\Gamma_{5}$ & $ O_{yz}=T^{\beta}_{y}T^{\beta}_{z}$
& $O_{zx}=T^{\beta}_{z}T^{\beta}_{x}$
& $O_{xy}=T^{\beta}_{x}T^{\beta}_{y}$ \\
$\Gamma_{3}$ & $O_{u}=\frac{1}{\sqrt{6}}(-T^{\beta}_{x}T^{\beta}_{x}
                            -T^{\beta}_{y}T^{\beta}_{y}
                           +2T^{\beta}_{z}T^{\beta}_{z})$ 
& $O_{v}=\frac{1}{\sqrt{2}}(T^{\beta}_{x}T^{\beta}_{x}
                           -T^{\beta}_{y}T^{\beta}_{y})$ \\
\hline
\end{halftabular}
\medskip
\end{table}

\begin{table}[t]
\caption{The sign of  $-\I sg(\brho)\EX^{\I\BQ_{\nu}\brho}$ on 
eight O sites and the hyperfine field on O sites.
The column of $\BQ_{\nu}$ shows $-\I sg(\brho)\EX^{\I\BQ_{\nu}\brho}$. The last column shows $H_{\rm hf}$,  eq.(\ref{eq3.2})  except  for the factor $C_{1,2}$.
}
\label{t1}
\begin{halftabular}{@{\hspace{\tabcolsep}\extracolsep{\fill}}crrrr}
\hline
 site($\brho$) &$\BQ_{x}$&$\BQ_{y}$&$\BQ_{z}$& total\\ \hline
 $\frac{a}{2}(3,3,3)$&  1      &    1    &      1  & $2\bI\cdot\bH$
 \\
 $\frac{a}{2}(3,3,1)$& -1      &   -1    &      1  & $-2I_{z}H_{z}$
 \\
 $\frac{a}{2}(1,1,3)$& -1      &   -1    &      1  & $-2I_{z}H_{z}$ 
 \\
 $\frac{a}{2}(1,1,1)$& 1        &   1    &      1  & $2\bI\cdot\bH$
 \\
 $\frac{a}{2}(3,1,3)$& -1        &   1    &   -1  &  $-2I_{y}H_{y}$
 \\
 $\frac{a}{2}(3,1,1)$& 1        &  -1    &     -1  & $ -2I_{x}H_{x}$
 \\
 $\frac{a}{2}(1,3,3)$& 1        &  -1    &     -1  & $ -2I_{x}H_{x}$
 \\
 $\frac{a}{2}(1,3,1)$& -1      &   1    &     -1  &  $-2I_{y}H_{y}$
 \\ 
\hline
\end{halftabular}
\medskip
\end{table}

\begin{table}[t]
\caption{Quadrupolar field on eight O sites. 
$O_{u}$ and $O_{v}$ are the quadrupolar operators of O nucleus.
The last column shows the result for 
$O_{yz}=O_{zx}=O_{xy}\equiv O_{\Gamma_{5}}$
}
\label{t2}
\begin{halftabular}{@{\hspace{\tabcolsep}\extracolsep{\fill}}crrc}
\hline
 site($\brho$) & $O_{u}$ & $O_{v}$ & total\\ \hline
 $\frac{a}{2}(3,3,3)$
& $\frac{1}{2}O_{yz}-O_{xy}+\frac{1}{2}O_{zx}$ 
& $\frac{\sqrt{3}}{2}(-O_{yz}+O_{zx})$
& $0$ \\
 $\frac{a}{2}(3,3,1)$
& $-\frac{1}{2}O_{yz}-O_{xy}-\frac{1}{2}O_{zx}$ 
& $\frac{\sqrt{3}}{2}(O_{yz}-O_{zx})$
& $-2O_{u}O_{\Gamma_{5}}$ \\
 $\frac{a}{2}(1,1,3)$
& $-\frac{1}{2}O_{yz}-O_{xy}-\frac{1}{2}O_{zx}$ 
& $\frac{\sqrt{3}}{2}(O_{yz}-O_{zx})$
& $-2O_{u}O_{\Gamma_{5}}$ \\
 $\frac{a}{2}(1,1,1)$
& $\frac{1}{2}O_{yz}-O_{xy}+\frac{1}{2}O_{zx}$ 
& $\frac{\sqrt{3}}{2}(-O_{yz}+O_{zx})$
& $0$ \\
 $\frac{a}{2}(3,1,3)$
& $-\frac{1}{2}O_{yz}+O_{xy}+\frac{1}{2}O_{zx}$ 
& $\frac{\sqrt{3}}{2}(O_{yz}+O_{zx})$
& $(O_{u}+\sqrt{3}O_{v})O_{\Gamma_{5}}$ \\
 $\frac{a}{2}(3,1,1)$
& $\frac{1}{2}O_{yz}+O_{xy}-\frac{1}{2}O_{zx}$ 
& $\frac{\sqrt{3}}{2}(-O_{yz}-O_{zx})$
& $(O_{u}- \sqrt{3}O_{v}O_{\Gamma_{5}}$ \\
 $\frac{a}{2}(1,3,3)$
& $\frac{1}{2}O_{yz}+O_{xy}-\frac{1}{2}O_{zx}$ 
& $\frac{\sqrt{3}}{2}(-O_{yz}-O_{zx})$
& $(O_{u}-\sqrt{3}O_{v})O_{\Gamma_{5}}$ \\
 $\frac{a}{2}(1,3,1)$
& $-\frac{1}{2}O_{yz}+O_{xy}+\frac{1}{2}O_{zx}$ 
& $\frac{\sqrt{3}}{2}(-O_{yz}-O_{zx})$
& $(O_{u}+\sqrt{3}O_{v})O_{\Gamma_{5}}$ \\
\\
\hline
\end{halftabular}
\medskip
\end{table}

\section{Magnetic Field Effect in NpO$_{2}$
}
The applied magnetic field  induces  AFM
moments with the same wave vector.  
They  can be derived by using the Table III of ref. \citen{A10}:
\Beqa
J_{x}(\BR)
&=&\frac{b(\Gamma_{4},\Gamma_{5})}{\sqrt{2}}
\Big ( O_{xy}H_{y}\EX^{\I\BQ_{z}\BR}
+O_{zx}H_{z}\EX^{\I\BQ_{y}\BR} \Big) \nonumber \\
&=&\frac{b(\Gamma_{4},\Gamma_{5})}{\sqrt{2}}
 O_{\Gamma_{5}}
\Big(H_{y}\EX^{\I\BQ_{z}\BR}+H_{z}\EX^{\I\BQ_{y}\BR} \Big),  \\
J_{y}(\BR) 
&=& \frac{b(\Gamma_{4},\Gamma_{5})}{\sqrt{2}}
 O_{\Gamma_{5}}
\Big(H_{x}\EX^{\I\BQ_{z}\BR}+H_{z}\EX^{\I\BQ_{x}\BR} \Big), \\
J_{z}(\BR)
&=& \frac{b(\Gamma_{4},\Gamma_{5})}{\sqrt{2}}
 O_{\Gamma_{5}}
\Big(H_{x}\EX^{\I\BQ_{y}\BR}+H_{y}\EX^{\I\BQ_{x}\BR} \Big),
\label{eq3.1}
\Eeqa
where $b(\Gamma_{4},\Gamma_{5})$ is a proportionality constant.
Substituting this expression into  eq. (\ref{eq2.1}),  
we obtain the hyperfine field as 
\Beqa
H_{\rm hf}(\brho,c_{1,2}) 
  & = &
C_{1,2}\I sg(\brho)
\Big[
  I_{x}H_{x}
\Big(\EX^{\I\BQ_{y}\brho}+\EX^{\I\BQ_{z}\brho} \Big)
\nonumber \\
 & &
+ I_{y}H_{y}
\Big(\EX^{\I\BQ_{z}\brho}+\EX^{\I\BQ_{x}\brho} \Big)
+ I_{z}H_{z}
\Big(\EX^{\I\BQ_{x}\brho}+\EX^{\I\BQ_{y}\brho} \Big)
\Big],
\label{eq3.2}
\Eeqa
where the quantity $C_{1,2}$
is proportional to $c_{1,2}$ and the AFQ order parameter, i.e.  
$C_{1,2} \propto c_{1,2}O_{\Gamma_{5}}b(\Gamma_{4},\Gamma_{5})$.
This type of interaction appears through the coupling term with the 
constant $c_{1,2}$ in eq. (\ref{eq2.1}).
If one considers only the term of the type $\vec{I}\cdot\vec{J}$,
which has the  coupling constant $c_{1,1}$ in eq. 
(\ref{eq2.1}),
the interaction expressed by  eq. (\ref{eq3.2}) will not appear.
In addition, this interaction  originates from the coupling caused by
 the T$_{d}$ symmetry, such as $s_{z}c_{x}c_{y}$ term in eq.(\ref{eq2.1}).

At the $\brho_{1}=\frac{1}{2}(3a,3a,3a)$ site,
the factor
$
-\I sg(\brho_{1})
\EX^{\I\BQ_{\nu}\brho_{1}}
$ 
takes 1 for all $\nu$.
Therefore 
we have
$
H_{\rm hf}(\brho_{1},c_{1,2})=-2C_{1,2}
\vec{I}\cdot\vec{H}
$.
On the other hand,  at the site 
$\brho_{2}=\frac{1}{2}(3a,3a,a)$,
the factor takes $-1, -1$ and $1$, respectively, for
the terms having $\BQ_{x},\BQ_{y}$ and $ \BQ_{z}$.
Then we have 
$
H_{\rm hf}(\brho_{2},c_{1,2})=2C_{1,2}
I_{z}H_{z}
$.
Clearly the hyperfine field on the $\brho_{1}$ site  is  
different from that on the  $\brho_{2}$ site. 
We find in this way that 
the eight O sites 
split into four inequivalent sites
also in hyperfine interaction,
 as summarized  in Table \ref{t1}. 
When the magnetic field is given by $(H_x, H_y, H_z)$, four  types of the hyperfine field are induced on the O atoms:
$
2C_{1,2}(  H_{x},  H_{y},  H_{z})
$,
$
2C_{1,2}(     0,      0,  - H_{z})
$,
$
2C_{1,2}(     0,     - H_{y},   0)
$,
and
$
2C_{1,2}( - H_{x},     0,      0)
$.

For polycrystalline samples we have to take an average of the splitting over the direction. Then, we find that 
the first group does not show any broadening of the resonance, 
because
the hyperfine field is always induced in the direction of the 
applied field.
On the other hand, the hyperfine field usually deviates from the
direction of the field in the other three groups.
This will cause the broadening of 
the resonance spectrum with the line shape of the powder pattern of 
 the axial 
symmetry crystal.

Let us remember that the magnetic field usually induces the AFO moments
of the types, $T^{\beta}$ ( whose wave number vector is not equal to that of the original 
one) and $T_{xyz}$,  as shown in Table III of 
ref. \citen{A10}.
The hyperfine field on the O atoms is induced also through these moments as 
can be seen in  eq. (\ref{eq2.1}).
They give the hyperfine field of the type:
\Beqa
H_{\rm hf}(\brho,c_{1,3}) 
  & = &
-C_{1,3}\I sg(\brho)
\Big[
  I_{x}H_{x}
\Big(\EX^{\I\BQ_{z}\brho}-\EX^{\I\BQ_{y}\brho} \Big)
\nonumber \\
 & &
+ I_{y}H_{y}
\Big(\EX^{\I\BQ_{x}\brho}-\EX^{\I\BQ_{z}\brho} \Big)
+ I_{z}H_{z}
\Big(\EX^{\I\BQ_{y}\brho}-\EX^{\I\BQ_{x}\brho} \Big)
\Big],
\label{eq3.3-1}
\\
H_{\rm hf}(\brho,c_{1,4})
 & = &
C_{1,4}\I sg(\brho)
\Big(\EX^{\I\BQ_{x}\brho}I_{x}H_{x}
+\EX^{\I\BQ_{y}\brho}I_{y}H_{y}
+\EX^{\I\BQ_{z}\brho}I_{z}H_{z} \Big),
\label{eq3.3-2}
\Eeqa
where 
$C_{1,3} \propto c_{1,3}O_{\Gamma_{5}}$ and 
$C_{1,4} \propto c_{1,4}O_{\Gamma_{5}}$. 
The former  gives the hyperfine field
$
-2C_{1,3}(  0,        0,     0)
$,
$
-2C_{1,3}(    H_{x},  H_{y}, 0)
$,
$
-2C_{1,3}(    0,     H_{y},   H_{z})
$,
and 
$
-2C_{1,3}( H_{x},     0,      H_{z}), 
$
whereas the latter gives
$
C_{1,4}(  H_{x},       H_{y},   H_{z})
$,
$
C_{1,4}(   - H_{x}, - H_{y}, H_{z})
$,
$
C_{1,4}(    H_{x},   -  H_{y},  - H_{z})
$,
and
$
C_{1,4}( - H_{x},     H_{y},     - H_{z})
$.

To apply this theory to NpO$_2$, we have to sum up these 
three contributions. 
Even after the summation we still have a sharp line and a broad line;  
the shift of the sharp line and the 
field induced  part of the width of the broad line
have comparable 
magnitude;  they are  proportional to the strength of the 
magnetic field in low field region.

Our theory has already predicted 
that 
the quadrupolar splitting with uniaxial symmetry appears
on the non-cubic O site 
in the ordered phase of NpO$_2$.
Even at  zero field, this quadrupolar splitting exits.
By the application of the magnetic field in the AFO state, 
the AFQ is also induced as given in the
table \ref{t4}.
We can see that 
the induced quadrupolar moments of the $\Gamma_{5}$ type:
$
O_{yz}(\BR)=\frac{c(\Gamma_{5},\Gamma_{5})}{\sqrt{2}}
(  H_{z}T^{\beta}_{y}\EX^{\I\BQ_{y}\BR}
 - H_{y}T^{\beta}_{z}\EX^{\I\BQ_{z}\BR})
$,
$
O_{zx}(\BR)=\frac{c(\Gamma_{5},\Gamma_{5})}{\sqrt{2}}
(  H_{x}T^{\beta}_{z}\EX^{\I\BQ_{z}\BR}
 - H_{z}T^{\beta}_{x}\EX^{\I\BQ_{x}\BR})
$ and
$
O_{xy}(\BR)=\frac{c(\Gamma_{5},\Gamma_{5})}{\sqrt{2}}
(  H_{y}T^{\beta}_{x}\EX^{\I\BQ_{x}\BR}
 - H_{x}T^{\beta}_{y}\EX^{\I\BQ_{y}\BR})
$
cause the quadrupolar splitting.
Here 
$c(\Gamma_{5},\Gamma_{5})$
is the proportionality coefficient in Table \ref{t4}.
Substituting these results into eq. (\ref{eq2.2}), 
we obtain,
\Beqa
H_{\rm qq}(\brho,c_{2,5})
 & = &
 C_{2,5}\I sg(\brho)
\Big[
  O_{yz}(\brho)H_{x}
\Big( \EX^{\I\BQ_{z}\brho}
 -\EX^{\I\BQ_{y}\brho}\Big)
\nonumber \\
 & & 
+O_{zx}(\brho)H_{y}
\Big( \EX^{\I\BQ_{x}\brho}
 -\EX^{\I\BQ_{z}\brho}\Big)
 +O_{xy}(\brho)H_{z}
\Big( \EX^{\I\BQ_{y}\brho}
 -\EX^{\I\BQ_{x}\brho}\Big)
\Big],
\label{eq3.4}
\Eeqa
where 
$C_{2,5} \propto c_{2,5}c(\Gamma_{5},\Gamma_{5})T^{\beta}$.
The phase factor of this expression has the same form as that of 
eq. (\ref{eq3.3-1}); thus it will be canceled out on the cubic sites.
On the non-cubic sites, 
the quadrupolar interaction terms of the type
$C_{2,5}[O_{yz}H_{x}+O_{zx}H_{y}]$,
$C_{2,5}[O_{zx}H_{y}+O_{xy}H_{z}]$ and
$C_{2,5}[O_{xy}H_{z}+O_{yz}H_{x}]$
are expected.

The broadening due to the quadrupolar splitting is caused
at non-cubic sites when the average is taken over the direction  
for polycrystalline 
samples.
Even at  zero field, the broadening due to the quadrupolar splitting 
will exist.\cite{E1}
Thus, in the low magnetic field region, the broadening of the spectrum 
originating from the non-cubic site will be the sum of a constant term and the term proportional to a  
strength of the magnetic field.

The field-induced  AFO/AFM moments are proportional to both the magnetic 
field and the 
AFQ moment. Therefore 
they are proportional to the square
of the primary AFO moment. 
Another type of broadening due to the field induced 
hyperfine splitting 
is also expected.
The AFM and the AFO moments, which are directly  proportional
 to the primary AFO moment and the square of the 
magnetic field can be also induced.
For example, from the $\Gamma_{5}-2$ column of Table \ref{t5} and 
eq. (\ref{eq2.1}), we have the  interaction
\Beqa
H_{\rm hf}(\brho) 
  & = &
C_{1,2}\I sg(\brho) 
\Big[
  I_{x}H_{y}H_{z}
\Big(\EX^{\I\BQ_{y}\brho}+\EX^{\I\BQ_{z}\brho} \Big)
\nonumber \\
 & &
+ I_{y}H_{z}H_{x}
\Big(\EX^{\I\BQ_{z}\brho}+\EX^{\I\BQ_{x}\brho} \Big)
+ I_{z}H_{x}H_{y}
\Big(\EX^{\I\BQ_{x}\brho}+\EX^{\I\BQ_{y}\brho} \Big)
\Big],
\label{eq3.5}
\Eeqa
where $C_{1,2}  \propto c_{1,2}T^{\beta}$.
This term has the largest contribution to  the cubic site, and the splitting due to this term 
depends on the field  direction as  $H^{2}h_{x}h_{y}h_{z}$. 
After averaging over the direction, this  effect causes 
a broadening of spectrum of $O(H^{2})$. However, it still gives two contributions  with the intensity 
ratio 1: 3. 
The equations (\ref{eq3.2}) $\sim$ (\ref{eq3.3-2}) must contain  the 
$O(H^{2})$ terms coming from the change of the order parameters.

Finally we conclude that the shift of the sharp line, 
$\omega_{s}$ depend on the field strength as
$\omega_{s}= a^{(s)}_{s}H+ O(H^{2}) $.
The width of the sharp line,
$\Delta_{s}$ will be  given as a higher order function of $H$,
i.e. $\Delta_{s}=O(H^{2})$.
The width of the broad line,
$\Delta_{b}$, is given as 
$\Delta_{b}=c^{(w)}_{b}+a^{(w)}_{b}H+O(H^{2})$,
while the shift of the broad line, $\omega_{b}$  
is $\omega_{b}=a^{(s)}_{b}H+O(H^{2})$.
The constant $c^{(w)}_{b}$ is caused by the quadrupolar splitting.
If the hyperfine interaction is the main mechanism, 
the constants $a^{(s)}_{s}$, $a^{(w)}_{b}$, $a^{(s)}_{b}$ will have 
same order of magnitude.
$a^{(s)}_{b}$ and $a^{(w)}_{b}$ are related to the angle average of the 
axial symmetric sites.
These four constants are proportional to the AFQ moment; thus they must 
have almost the same temperature dependence.
If the magnetic field induced AFQ in the AFO state,
eq (\ref{eq3.4}),
 has appreciable contribution,
its effect will appear mainly in the $a^{(w)}_{b}$ term.
This contribution is directly proportional to the AFO order parameter.

\begin{table}[t]
\caption{Quadrupolar moment as a product of magnetic field 
and octupolar moment.}
\label{t4}
\begin{halftabular}{@{\hspace{\tabcolsep}\extracolsep{\fill}}ccc}
\hline
 Quadrupole moment & $\Gamma_{5}$ & $\Gamma_{2}$ \\ \hline
 $O_{yz}$
& $\frac{1}{\sqrt{2}}(- H_{y}T^{\beta}_{z}+ H_{z}T^{\beta}_{y}) $
& $ H_{x}T_{xyz}$ \\
 $O_{zx}$
& $\frac{1}{\sqrt{2}}(- H_{z}T^{\beta}_{x}+ H_{x}T^{\beta}_{z}) $
& $  H_{y}T_{xyz}$ \\
 $O_{xy}$
& $\frac{1}{\sqrt{2}}(- H_{x}T^{\beta}_{y}+ H_{y}T^{\beta}_{x}) $
& $  H_{z}T_{xyz}$ \\
 $O_{u}$
& $\frac{1}{\sqrt{2}}(- H_{x}T^{\beta}_{x}+ H_{y}T^{\beta}_{y}) $ \\
 $O_{v}$
& $\frac{1}{\sqrt{6}}(- H_{x}T^{\beta}_{x}-  H_{y}T^{\beta}_{y}
                      +2H_{z}T^{\beta}_{z})$ \\
\hline
\end{halftabular}
\medskip
\end{table}

\begin{table}[t]
\caption{$\Gamma_{4}$ type moment as a product of magnetic field 
and octupolar moment.}
\label{t5}
\begin{halftabular}{@{\hspace{\tabcolsep}\extracolsep{\fill}}cccc}
\hline
$\Gamma_{4}$ moment & $\Gamma_{5}-1$ & $\Gamma_{5}-2$ & $\Gamma_{2}$\\ \hline
 $T_{x}^{\alpha}$
& $-\frac{1}{\sqrt{2}}(H^{2}_{y}- H^{2}_{z})T^{\beta}_{x}$
& $ H_{x}(H_{y}T^{\beta}_{y}- H_{z}T^{\beta}_{z})$
& $\sqrt{2}H_{y}H_{z}T_{xyz}$ \\
 $T_{y}^{\alpha}$
& $-\frac{1}{\sqrt{2}}(H^{2}_{z}- H^{2}_{x})T^{\beta}_{y}$
& $ H_{y}(H_{z}T^{\beta}_{z}- H_{x}T^{\beta}_{x})$
& $\sqrt{2}H_{z}H_{x}T_{xyz}$ \\
 $T_{z}^{\alpha}$ 
& $-\frac{1}{\sqrt{2}}(H^{2}_{x}- H^{2}_{y})T^{\beta}_{z}$
& $ H_{z}(H_{x}T^{\beta}_{x}- H_{y}T^{\beta}_{y})$
& $\sqrt{2}H_{x}H_{y}T_{xyz}$ \\
\hline
\end{halftabular}
\medskip
\end{table}

\begin{table}[t]
\caption{$\Gamma_{5}$ type moment as a product of magnetic field 
and octupolar moment.}
\label{t6}
\begin{halftabular}{@{\hspace{\tabcolsep}\extracolsep{\fill}}cccc}
\hline
$\Gamma_{5}$ moment & $\Gamma_{5}-1$ & $\Gamma_{5}-2$& $\Gamma_{5}-3$ \\ \hline
 $T_{x}^{\beta}$
& $-\frac{1}{\sqrt{6}}(2H^{2}_{x}-H^{2}_{y}-H^{2}_{z})T^{\beta}_{x}$
& $- H_{x}(H_{y}T^{\beta}_{y}+H_{z}T^{\beta}_{z})$
& $\frac{1}{\sqrt{3}}H^{2}T^{\beta}_{x}$ \\
 $T_{y}^{\beta}$
& $-\frac{1}{\sqrt{6}}(2H^{2}_{y}- H^{2}_{z}-H^{2}_{x})T^{\beta}_{y}$
& $- H_{y}(H_{z}T^{\beta}_{z}+H_{x}T^{\beta}_{x})$
& $\frac{1}{\sqrt{3}}H^{2}T^{\beta}_{y}$ \\
 $T_{z}^{\beta}$
& $-\frac{1}{\sqrt{6}}(2H^{2}_{z}- H^{2}_{x}-H^{2}_{y})T^{\beta}_{z}$
& $- H_{z}(H_{x}T^{\beta}_{x}+H_{y}T^{\beta}_{y})$
& $\frac{1}{\sqrt{3}}H^{2}T^{\beta}_{z}$ \\
\hline
\end{halftabular}
\medskip
\end{table}

\section{Hyperfine Interaction with Multipolar Moments in CeB$_{6}$ and Ce$_{x}$La$_{1-x}$B$_{6}$}

CeB$_{6}$ has the crystal structure of CaB$_{6}$ type. 
The Ce ions form the s.c. lattice.
Let us denote the B ion site
($\frac{a}{2},\frac{a}{2},\pm ua$)  as 
$\brho_{z,\pm}=(0,0,\pm\frac{1}{2})$,
where $u$ is a parameter for the position of B  atom and $a$ is the lattice constant of the s.c. 
lattice. 
If we follow the method described in Appendix C,  
the interaction of the nuclear spin $\BI$ of B atom  with the multipolar moments of Ce ions 
is given by
\Beqa
H_{\rm hf}(\brho_{z,\pm})
 & = &
\frac{4}{\sqrt{4}}\EX^{\I \frac{q_{x}+q_{y}+q_{z}}{2} a}
\EX^{\I\Bq\brho_{z,\pm}}
\Big[ I_{z}(\brho_{z,\pm})
\Big\{ c_{1,1}J_{z}(\Bq)c_{x}c_{y}
\nonumber \\
  &  &
+\I sg(\brho_{z,\pm}) 
\frac{c_{1,2}}{\sqrt{2}}
\Big(  J_{x}(\Bq)s_{x}c_{y}
 +J_{y}(\Bq)c_{x}s_{y}
\Big) 
+\I sg(\brho_{z,\pm}) 
\frac{c_{1,3}}{\sqrt{2}}
\Big( T^{\beta}_{x}(\Bq) s_{x}c_{y} 
 -T^{\beta}_{y}(\Bq) c_{x}s_{y}
\Big) 
\nonumber \\ 
  &  &
- c_{1,4}
T_{xyz}(\Bq)s_{x}s_{y} 
\Big\}
\nonumber \\
  & +  &
I_{x}(\brho_{z,\pm})
\Big\{
 c_{2,1}J_{x}(\Bq)c_{x}c_{y}
+\I sg(\brho_{z,\pm})
 c_{2,2}J_{z}(\Bq)s_{x}c_{y}
-c_{2,3}J_{y}(\Bq)s_{x}s_{y}
\nonumber \\
  &  &
+c_{2,4}T^{\beta}_{x}(\Bq)c_{x}c_{y}
+c_{2,5}T^{\beta}_{y}(\Bq)s_{x}s_{y}
+\I sg(\brho_{z,\pm})
c_{2,6}T^{\beta}_{z}(\Bq)s_{x}c_{y}
+ \I sg(\brho_{z,\pm})
c_{2,7}T_{xyz}(\Bq) c_{x}s_{y}
\Big\}
\nonumber \\
  & +  &
I_{y}(\brho_{z,\pm})
\Big\{
 c_{2,1}J_{y}(\Bq)c_{x}c_{y}
+\I sg(\brho_{z,\pm})
c_{2,2}J_{z}(\Bq) s_{y}c_{x}
-c_{2,3}J_{x}(\Bq)s_{x}s_{y}
\nonumber \\
  &  &
+c_{2,4}T^{\beta}_{y}(\Bq)c_{x}c_{y}
-c_{2,5}T^{\beta}_{x}(\Bq)s_{x}s_{y}
+\I sg(\brho_{z,\pm})
c_{2,6}T^{\beta}_{z}(\Bq) s_{y}c_{x}
\nonumber \\
  &  &
+\I sg(\brho_{z,\pm})
c_{2,7}T_{xyz}(\Bq) c_{y}s_{x}
\Big\}
\Big] .
\label{eq4.1}
\Eeqa
Here $c_{i,j}$'s  are coupling constants, 
and 
$ sg(\brho_{z,\pm})=\pm1$.
The quantity $c_{\nu}$($s_{\nu}$) represents $\cos(q_{\nu}/2)$
($\sin( q_{\nu}/2)$). 
The result (\ref{eq4.1}) agrees with the result given  in ref. \citen{A8}.

The hyperfine interaction of the B nuclei  at the 
$(\pm ua, \frac{a}{2},\frac{a}{2})$ site, which is denoted henceforth as 
$\brho_{x,\pm}=(\pm\frac{1}{2},0,0)$, 
is obtained by 90$^{\circ}$ degree rotation around the $y$-axis:
$
(I_{x},I_{y},I_{z}) \rightarrow 
(-I_{z},I_{y},I_{x})$,
$
(T^{\beta}_{x},-T^{\beta}_{y},T^{\beta}_{z}) 
\rightarrow 
(T^{\beta}_{z},-T^{\beta}_{y},-T^{\beta}_{x}) 
$,
\hspace{0.2cm}
$T_{xyz} 
\rightarrow
-T_{xyz}
$.
At the same time the wave vector is changed as 
$(q_{x},q_{y},q_{z}) \rightarrow 
(-q_{z},q_{y},q_{x})$
except  for the factor
in the function
$\EX^{\I\frac{q_{x}+q_{y}+q_{z}}{2}}\EX^{\I\Bq\brho_{x,\pm}}$.
The dipole operator $(J_{x},J_{y},J_{z})$ 
follows the same transformation
to that of the operator $(I_{x},I_{y},I_{z})$.
Similarly, the hyperfine interaction for the B on the 
$(\frac{a}{2},  \pm ua,  \frac{a}{2} )$ site, which is denoted as
$\brho_{y,\pm}=(0,\pm\frac{1}{2},0)$, is
obtained by a rotation around the $x$-axis:
$(I_{x},I_{y},I_{z}) \rightarrow 
(I_{x},-I_{z},I_{y})$,
$
(T^{\beta}_{x},-T^{\beta}_{y},T^{\beta}_{z}) 
\rightarrow 
(-T^{\beta}_{x},T^{\beta}_{z},-T^{\beta}_{y}) 
$,
$T_{xyz} 
\rightarrow
-T_{xyz}
$,
and
$(q_{x},q_{y},q_{z}) \rightarrow 
(q_{x},-q_{z},q_{x})$.

If  we assume the 
$\BQ =\pi(1,1,1)$ order 
as done in refs. (\citen{A6}) and (\citen{A14}),
the quantities $s_{\nu}$ and $c_{\nu}$ are reduced to 
$s_{x}=s_{y}=s_{z}=1$, and 
$c_{x}=c_{y}=c_{z}=0$.  Then  the hyperfine field  is expressed as follows: 
\Beqa
H_{\rm hf}(\brho_{z,\pm})
 &= &
\pm
\Big[ 
I_{z}(\brho_{z,\pm})
( -C_{1,4}T_{xyz}(\BQ) )
\nonumber \\
& &
+I_{x}(\brho_{z,\pm})
\Big(
-C_{2,3}J_{y}(\BQ)+C_{2,5}T^{\beta}_{y}(\BQ)
\Big)
\nonumber \\
& &
+I_{y}(\brho_{z,\pm})
\Big(
-C_{2,3}J_{x}(\BQ)-C_{2,5}T^{\beta}_{x}(\BQ)
\Big)
\nonumber \\
 & &
+I_{z}
(\brho_{z,\pm})
(-C_{1,1}J_{z}(0))
\nonumber \\
 & &
+I_{x}(\brho_{z,\pm})
\Big(
C_{2,1}J_{x}(0)+C_{2,4}T^{\beta}_{x}(0)
\Big)
+I_{y}(\brho_{z, \pm})
\Big(
C_{2,1}J_{y}(0)+C_{2,4}T^{\beta}_{y}(0)
\Big)
\Big]\ .
\label{eq4.2}
\Eeqa
Here $T^{\beta}_{\nu}(\BQ)$, $J_{\nu}(\BQ)$ and $T_{xyz}(\BQ)$ are the 
thermal average of the  ordering parameters with wave vector $\BQ$.
The quantities for $\BQ=0$ are the uniform components.
The factor $C_{i,j}$ is a  constant proportional to $c_{i,j}$.
Even in the uniform term we have the hyperfine interaction due to the 
octupolar moment $T^{\beta}_{\nu}(0)$.
Such term will be induced by the application of the magnetic field 
under the existence of the uniform component of quadrupolar moment.
It is given by a different combination of the quadrupolar moment and the
magnetic field from that of $J_{\nu,0}$ term\cite{A10}.
Hereafter we neglect the $\BQ=0$ terms.
The hyperfine interaction for $\brho_{x,\pm}$
and  $\brho_{y,\pm}$ site is given ,
\Beqa
H_{\rm hf}(\brho_{x,\pm})
 &= &
\pm
\Big[
I_{x}(\brho_{x,\pm})
(-C_{1,4}T_{xyz}(\BQ))
\nonumber \\
& &
+I_{z}(\brho_{x,\pm})
\Big\{
-C_{2,3}J_{y}(\BQ)-C_{2,5}T^{\beta}_{y}(\BQ)
\Big\}
\nonumber \\
& & 
+I_{y}(\brho_{x,\pm})
\Big\{
-C_{2,3}J_{z}(\BQ)+C_{2,5}T^{\beta}_{z}(\BQ)
\Big\}
\Big]
,
\label{eq4.3}
\Eeqa

\Beqa
H_{\rm hf}(\brho_{y,\pm})
 &= &
\pm
\Big[ 
I_{y}(\brho_{y,\pm})
(-C_{1,4}T_{xyz}(\BQ))
\nonumber \\
& & 
+I_{x}(\brho_{y,\pm})
\Big\{
-C_{2,3}J_{z}(\BQ)-C_{2,5}T^{\beta}_{z}(\BQ)
\Big\}
\nonumber \\
& & 
+I_{z}(\brho_{y,\pm})
\Big\{
-C_{2,3}J_{x}(\BQ)+C_{2,5}T^{\beta}_{x}(\BQ)
\Big\}
\Big]
.
\label{eq4.4}
\Eeqa

If  the ordering wave vector 
$
\Bq
$
satisfies
$\cos(q_{y}/2) \neq \cos( q_{z}/2)$, a  mixing
of 
$J_{x} (\Bq)$ with $T^{\beta}_{x}(\Bq)$
through the nearest site interaction on the s.c. lattice\cite{A10,A16}
will be caused.
In addition, the ordering with 
$ \sin( q_{\nu}) \neq 0$ 
causes mixing of dipole and 
$T^{\beta}$
type moment through the next nearest site interaction\cite{A10}.
The ordering of the wave vector with 
$\BQ=\pi(1,1,1)$
seems to be a plausible candidate
when one considers the pure spontaneous AFO.
The n.n. interaction of $\Gamma_{5}$ 
moment will have comparable magnitude to that of the quadrupolar and 
magnetic interaction\cite{A16}. 

\section{Dependence of Hyperfine Splitting on the Field Direction
 in Ce$_{1-x}$La$_{x}$B$_{6}$
}

Let us assume that  the ordering  with
$\BQ=\pi(1,1,1)$ occurs. Then,  the hyperfine field splitting $\Delta(z)$ of B-ion pair at 
$(\frac{a}{2},\frac{a}{2},\pm ua)$ is given as
\Beqa
\Delta(z) & = &
-2C_{1,4}h_{z} T_{xyz}(\BQ)
+2C_{2,5}(h_{x} T^{\beta}_{y}(\BQ) - h_{y} T^{\beta}_{x}(\BQ))
\nonumber \\ 
 & &
-2C_{2,3}(h_{x} J_{y}(\BQ) + h_{y} J_{x}(\BQ)),
\label{eq5.1}
\Eeqa
where $
(h_{x},h_{y},h_{z})$ denotes the direction of the magnetic field.
Henceforth we consider the ordering of three types: 
\par
\parindent 0pt 
(i) $ T^{\beta}_{x}(\BQ)= T^{\beta}_{y}(\BQ) = T^{\beta}_{z}(\BQ)
\equiv \frac{1}{\sqrt{3}}T^{\beta}_{111}$,
\par
(ii)  $ T^{\beta}_{x}(\BQ) =- T^{\beta}_{y}(\BQ)
\equiv \frac{1}{\sqrt{2}}T^{\beta}_{1\bar{1}0},
 T^{\beta}_{z}(\BQ) = 0$, 
 \par
(iii) $ T^{\beta}_{x}(\BQ) = T^{\beta}_{y}(\BQ)
\equiv -\frac{1}{\sqrt{6}}T^{\beta}_{110}$, and 
$ T^{\beta}_{z}(\BQ)
\equiv \frac{2}{\sqrt{6}}T^{\beta}_{\bar{1}\bar{1}2}$.
\par
\parindent 10pt

In the case of
$ T^{\beta}_{x}(\BQ)=
 T^{\beta}_{y}(\BQ)$ ((i) and (iii)),
the expression (\ref{eq5.1}) depends on the direction of the magnetic field as
$
\sin\TH\sin(\phi-\frac{\pi}{4})
$.
Here the direction of the magnetic field is expressed as 
$(\sin\TH\cos\phi,\sin\TH\sin\phi,\cos\TH)$.
$\Delta(z)$ vanishes at $\phi =\pi/4$, 
and has linear dependence, as $\Delta(z) \propto \phi-\frac{\pi}{4}$, 
in the vicinity of  $\phi=\frac{\pi}{4}$.
In the case of 
$ T^{\beta}_{x}(\BQ)=
- T^{\beta}_{y}(\BQ) $ (case (ii)),
$\Delta(z)$ depends on the direction as 
$\sin\TH\sin(\phi+\frac{\pi}{4})$, showing the maximum at $\phi=\pi/4$.

When the magnetic field is applied, the  order 
parameters are modified. 
The lowest-order effect is given by the product of the 
square of the magnetic field and the linear term of the order 
parameters, because they are the lowest order expression with
non-uniform and time reversal odd in the case of AFO ordering.
In Table \ref{t5}, we show the induced  $\Gamma_{4}$ AFO moment 
as a product of the $\Gamma_{5}$ AFO moment and the magnetic field.
This will also be proportional to the dipole  moment.
Substituting this into the third term of eq. (\ref{eq5.1}),
we have 
\Beqa
 h_{x}J_{y}(\BQ)+ h_{y}J_{x}(\BQ)
\hspace{10cm}
\nonumber \\
 =   (h_{x}-h_{y})
\Big[
\frac{a_1(\Gamma_{4},\Gamma_{5})}{\sqrt{2}}(h^{2}_{x}
+h_{x}h_{y}+h^{2}_{y}-h^{2}_{z})
-a_2(\Gamma_{4},\Gamma_{5})h_{x}h_{y}\Big] H^{2}T^{\beta}_{\parallel},
\label{eq5.2}
\Eeqa
where 
$
a_1(\Gamma_{4},\Gamma_{5})
$ and 
$
a_2(\Gamma_{4},\Gamma_{5})
$
are proportionality constants.
Here we introduced 
$ T^{\beta}_{x}(\BQ)= T^{\beta}_{y}(\BQ) \equiv T^{\beta}_{\parallel}$.
The terms proportional to the order parameter
$ T^{\beta}_{z}(\BQ)$ do not appear even when it exits.
Equation  (\ref{eq5.2}) is also proportional to $h_{x}-h_{y}$, thus 
the linear dependence on $\phi-\pi/4$ is not changed.
The induced moments of $\Gamma_{5}$ type can be derived also  by using the 
Table \ref{t6}, but the conclusion is unchanged.

For the order of  the type
$ J_{x}(\BQ) =- J_{y}(\BQ)$,
the same $\phi$ dependence 
as  that of $T^{\beta}_{x}(\BQ)=T^{\beta}_{y}(\BQ)$ 
is expected as seen from eq. (\ref{eq5.1}).
The field dependence is also similar as one can check  easily by using
the results of Tables \ref{t7} and \ref{t8}. 
However, complementary measurements of the splitting for pairs 
on the other axis, and/or 
detailed calculations of the direction dependence 
based on the microscopic model 
will distinguish them.
If the  
$
T_{xyz}(\BQ)
$
order is realized, the dependence of $\Delta(z)$  on the field direction is unique. 
This has been already noted in ref. \citen{A8}.

The splitting of the NMR of the B ion pair at 
$(\pm ua, \frac{a}{2},\frac{a}{2})$, 
$\Delta(x)$, is given as 
\Beqa
\Delta(x)
  & = &
-2C_{1,4}h_{x} T^{\beta}_{z}(\BQ)
+2C_{2,5}\Big(h_{z} T^{\beta}_{y}(\BQ)  - h_{y} T^{\beta}_{z}(\BQ) \Big)
\nonumber \\ 
 & &
-2C_{2,3}\Big(h_{z} J_{y}(\BQ)+ h_{y} J_{z}(\BQ)\Big).
\label{eq5.3}
\Eeqa
The splitting of the pair at $(\frac{a}{2},\pm ua, \frac{a}{2})$, 
$\Delta(y)$, is similarly given by using 
eq. (\ref{eq4.4}).
When  the octupolar ordering follows 
$ T^{\beta}_{z}(\BQ)=
 T^{\beta}_{y}(\BQ)$ (type (i)),
this expression has a node  
when $h_{z}=h_{y}$
and linear dependence 
around the direction.
The ordering of the type
 $ J_{z}(\BQ) =- J_{y}(\BQ) $
will also give the same direction dependence.
However, 
they can be distinguished when referring to the experimental
result of $\Delta(z)$.
The ordering of the type
$
 J_{x}(\BQ)= J_{z}(\BQ) =- J_{y}(\BQ)
$ 
does not have symmetry around the $[111]$ direction.
Such ordering  usually does not appear.
But the ordering of the type (i) can appear.
Moreover, the splitting 
$\Delta(y)$
has different direction dependence 
between them.
\begin{table}[t]
\caption{$\Gamma_{4}$ type moment as a product of magnetic field 
and dipole moment}
\label{t7}
\begin{halftabular}{@{\hspace{\tabcolsep}\extracolsep{\fill}}cccc}
\hline
$\Gamma_{4}$ moment & $\Gamma_{4}-1$ & $\Gamma_{4}-2$ & $\Gamma_{4}-3$\\ \hline
 $T_{x}^{\alpha}$
& $-\frac{1}{\sqrt{6}}(2H^{2}_{x}-H^{2}_{y}-H^{2}_{z})J_{x}$
& $- H_{x}(H_{y}J_{y}+H_{z}J_{z})$
& $\frac{1}{\sqrt{3}}H^{2}J_{x}$ \\
 $T_{y}^{\alpha}$
& $-\frac{1}{\sqrt{6}}(2H^{2}_{y}-H^{2}_{z}-H^{2}_{x})J_{y}$
& $- H_{y}(H_{z}J_{z}+H_{x}J_{x})$
& $\frac{1}{\sqrt{3}}H^{2}J_{y}$ \\
 $T_{z}^{\alpha}$
& $-\frac{1}{\sqrt{6}}(2H^{2}_{z}-H^{2}_{x}-H^{2}_{y})J_{z}$
& $- H_{z}(H_{x}J_{x}+H_{y}J_{y})$
& $\frac{1}{\sqrt{3}}H^{2}J_{z}$ \\
\hline
\end{halftabular}
\medskip
\end{table}
\begin{table}[t]
\caption{$\Gamma_{5}$ type moment as a product of magnetic field 
and dipole moment.}
\label{t8}
\begin{halftabular}{@{\hspace{\tabcolsep}\extracolsep{\fill}}cccc}
\hline
$\Gamma_{5}$ moment & $\Gamma_{4}-1$ & $\Gamma_{4}-2$ &    \\ \hline
 $T_{x}^{\beta}$
& $\frac{1}{\sqrt{2}}(H^{2}_{y}-  H^{2}_{z})J_{x}$
& $ H_{x}(H_{y}J_{y}- H_{z}J_{z})$
&  \\
 $T_{y}^{\beta}$
& $\frac{1}{\sqrt{2}}(H^{2}_{z}-  H^{2}_{x})J_{y}$
& $ H_{y}(H_{z}J_{z}- H_{x}J_{x})$
&  \\
 $T_{z}^{\beta}$
& $\frac{1}{\sqrt{2}}(H^{2}_{x}-  H^{2}_{y})J_{z}$
& $ H_{z}(H_{x}J_{x}- H_{y}J_{y})$
&  \\
\hline
\end{halftabular}
\medskip
\end{table}

In the ordering of type (ii), the splitting
$\Delta(x)$ has the largest value for the field in the 
$z$ direction.
In the ordering of type (iii), the splitting 
$\Delta(x)$ becomes zero
for $h_{z}/h_{y}=-2$, and takes a maximum at 
$ h_{y}/h_{z}=2$.

\begin{table}[t]
\caption{$\Gamma_{2}$($\Gamma_{1}$) type moment as a product of 
magnetic field and the $\Gamma_{4}$($\Gamma_{5}$) moment}
\label{t9}
\begin{halftabular}{@{\hspace{\tabcolsep}\extracolsep{\fill}}cc}
\hline
    $\Gamma_{4}$ & $\Gamma_{5}$ \\ \hline
 $T_{xyz}
= -\sqrt{\frac{2}{3}}(H_{y}H_{z}J_{x}+H_{z}H_{x}J_{y}+H_{x}H_{y}J_{z})
 $
& 
 $A_{1}=\sqrt{\frac{2}{3}}(H_{y}H_{z}T^{\beta}_{x}
                         + H_{z}H_{x}T^{\beta}_{y}
                         + H_{y}H_{z}T^{\beta}_{z})
$
 \\
\hline
\end{halftabular}
\medskip
\end{table}

\section{Summary}

We have presented a new approach to derive the hyperfine interaction 
between the nuclear spin
of ligand atom and multipolar moments of magnetic ions.

The most important part of this paper is the analysis on $^{17}$O NMR in NpO$_2$, which has been reported 
recently by Tokunaga et al. {\it et. al}\cite{A15}.  We have studied the $^{17}$O NMR spectrum for the 
triple $\bq$ order of primary AFO and secondary AFQ, which was proposed by Paix\~{a}o {\it et. al}.\cite{A4}. 
 At zero field the triple $\bq$ AFO does not produce any hyperfine field on the O sites. 
However, the secondary triple $\bq$ AFQ order causes 4 different quadrupolar fields on the eight O sites 
in the fcc cube: a pair of  cubic sites (with zero quadrupolar field) and 3 pairs of  uniaxial-symmetry sites 
(the principal axis:  $x, y$      or $z$ axis). 
When the magnetic field is applied,  the AFM/AFO is induced
 in cooperation with the preexisting triple $\bq$ AFQ order. As a result, 
the NMR spectrum from the cubic site gives a sharp line with a shift proportional to the strength
of the magnetic field because the hyperfine field is isotropic.  On the other hand, 
the spectrum from three kinds of non-cubic sites gives the same  shape for polycrystalline samples. Namely, 
it has a broadening due to the quadrupolar splitting and the magnetic
field induced part caused by  the anisotropic hyperfine field.  
The experimental observation by Tokunaga et al. is summarized as follows:
\par
\parindent 0pt
(I) splitting to sharp and broad lines with the ratio 1:3 in the ordered phase, 
\par
(II) the magnetic field dependence of the shift of the sharp line, 
\par
(III) the magnetic field dependence of the width of the broad line,
\par
(IV) relation between the  shift and the magnetic field induced part of the width. 
\par
Those features are  consistently explained by our scenario.
The present calculation and the experimental result in \citen{A15}
strongly support to the triple $\bq$ ordering model of NpO$_2$. 
\par
\parindent 10pt
We have found theoretically that in the AFO state there are unique coupling terms such as 
the one causing 
the quadrupolar splitting proportional 
to the magnetic field and/or the hyperfine field splitting proportional
to the square of the magnetic field.
The NMR experiment for single crystals is highly desirable. 
On the theoretical side a microscopic calculation including the effect of finite  magnetic field is desirable.  
In addition, although our phenomenological approach is certainly useful, a microscopic calculation of the transferred hyperfine interaction based on a 
microscopic model\cite{D5} is 
also desired.
Such microscopic calculations may  modify quantitatively  the results of the present paper,
but we believe they will not change our conclusion drastically.

In the second part we have 
derived the hyperfine interaction between the B nuclear spin 
and the multipolar moments  of Ce ion in CeB$_{6}$ and Ce$_{1-x}$La$_{x}$B$_{6}$. 
It has been applied to discuss a possibility to identify  the  octupolar 
order parameter, which was proposed for phase IV of Ce$_{1-x}$La$_{x}$B$_{6}$. We assumed 
the pure octupolar order of the 
$\Gamma_{5}$($T^{\beta}$) type with wave vector
$\BQ=\pi(1,1,1)$ 
under uniaxial stress of the $[1,1,1]$ direction.
If it is realized  in  Ce$_{1-x}$La$_{x}$B$_{6}$, the hyperfine field splitting of B nuclear spin 
 should show characteristic features dependent  on the type of order.
For example, if  the order of type
$T^{\beta}_{x}=T^{\beta}_{y}=T^{\beta}_{z}$ occurs, the splitting 
$\Delta(z)$ of NMR vanishes when the direction of the
magnetic field crosses the $(1\bar{1}0)$ plane.
Therefore, in principle, the octupolar ordering can be identified from the direction dependence 
of  the hyperfine splitting.  Magishi {\it et. al.} have made an NMR experiment on  the 
phase IV\cite{A11}.
Though the observed lines are sharp enough  in phase I (normal phase) , they overlap and become broad in phase IV. 
They speculated from the experiment that an AF magnetic ordering of incommensurate
wave vector is realized.  
The NMR experiment under a strong uniaxial stress along $(1,1,1)$
is desirable.  Although  the pure AFO ordering has been assumed for phase IV, 
 Ce$_{1-x}$La$_{x}$B$_{6}$ is actually a  disordered system so that  the AFO is accompanied inevitably by  AFM clusters because of low local symmetry.  This effect should show up in the neutron scattering experiment. In this connection we note that strange insensibility of Ce ions to the magnetic ordering in 
Nd$_{1-x}$Ce$_{x}$B$_{6}$ may be related to this problem.\cite{D2,D3}.

\section*{Acknowledgments}
The authors would like to express their sincere thanks for 
enlightening discussions  with S. Kambe on their NMR experiments. 
They also  would like to thank T. Morie, T. Sakakibara and Y. Tokunaga 
for sending us  preprints
before publication.
They thank  H. Kusunose and  Y. Kuramoto
 for useful discussions
on the octupolar interaction, 
and M. Kawakami for helpful discussions on their NMR experiments.
 They are also indebted to M. Sera for his continuous encouragement. 
The numerical computation was partly performed in the Computer Center
of Tohoku University and the Supercomputer Center of Institute for Solid
 State Physics (University of Tokyo).
The symmetrized bases are generated by using TSPACE pack which was coded
by A. Yanase (Tokyo, Shokabo, 1995).
This work was partly supported by JSPS and MEXT, KAKENHI(No.14340108) 
and (No.15034213).

\vfill\eject

\appendix

\section{Derivation of Invariant Coupling Form of Hyperfine interaction
in NpO$_{2}$
}

Let us denote the O site ($\frac{3a}{2},\frac{3a}{2},\frac{3a}{2}$) as 
$\brho_{1}$.
We
consider the invariant coupling form between the nuclear spin
$\BI$ 
of $^{17}$O on this site 
and multipolar 
moments of f-electrons of Np ions at four nearest neighbor sites,
$(2a,2a,2a)$, $(2a,a,a)$, $(a,2a,a)$ and $(a,a,2a)$
of the cube of the f.c.c. lattice. 
For simplicity we denote these Np ions as
$
(111), (1\bar{1}\bar{1}), (\bar{1}1\bar{1}),
$ 
and
$(\bar{1}\bar{1}1)$, respectively.
These ions have the T$_{\rm d}$ symmetry around the O atom.
In a previous paper, the invariant form is derived by considering 
the
pair interaction and rotating the pair by the 
symmetry operation\cite{A8}.
This method  certainly gives the correct interaction form, but 
as noted in ref. \citen{A10}, the calculation becomes 
relatively easier if one 
uses the symmetrized combination of multipolar operators.
The nuclear dipole  operators $I_{x}$, $I_{y}$ and $I_{z}$
form the 3-dimensional T$_{1}$
representation of T$_{\rm d}$\cite{G1}.
Following the method to make the symmetrized molecular orbital,
we can construct combinations of $\Gamma_{4}$, and $\Gamma_{5}$ and 
$\Gamma_{2}$  type 
odd power operators of the 
Np ions, which have T$_{1}$ representation.

Then, the invariant form of the hyperfine interaction of bilinear type is given as follows:
\Beqa
H_{\rm hf}(\brho_{1})
 & = &
I_{x}(\brho_{1})
\Big[
c_{1,1}\frac{1}{\sqrt{4}}
\Big\{ (J_{x})_{(111)}
+(J_{x})_{(1\bar{1}\bar{1})}
+(J_{x})_{(\bar{1}1\bar{1})}
+(J_{x})_{(\bar{1}\bar{1}1)}
\Big\}
\nonumber \\
& &
+c_{1,2}\frac{1}{\sqrt{8}}
\Big\{
 ( J_{y} +J_{z}))_{(111)}
+(-J_{y} -J_{z})_{(1\bar{1}\bar{1})}
+(-J_{y} +J_{z})_{(\bar{1}1\bar{1})}
+(J_{x}  -J_{z})_{(\bar{1}\bar{1}1)}
\Big\}
\nonumber \\
& &
+c_{1,3}\frac{1}{\sqrt{8}}
\Big\{
 ( T^{\beta}_{y} -T^{\beta}_{z}))_{(111)}
+(-T^{\beta}_{y} +T^{\beta}_{z})_{(1\bar{1}\bar{1})}
+(-T^{\beta}_{y} -T^{\beta}_{z})_{(\bar{1}1\bar{1})}
+( T^{\beta}_{y} +T^{\beta}_{z})_{(\bar{1}\bar{1}1)}
\Big\}
\nonumber \\
& &
+c_{1,4}\frac{1}{\sqrt{4}}
\Big\{
 ( T_{xyz})_{(111)}
+( T_{xyz})_{(1\bar{1}\bar{1})}
+(-T_{xyz})_{(\bar{1}1\bar{1})}
+(-T_{xyz})_{(\bar{1}\bar{1}1)}
\Big\}
\Big]
\nonumber \\
 & &
+ ({\rm cyclic}\ {\rm permutation}\ {\rm of}\  x,   y  \ {\rm and}\  z)\ .
\label{eqA.2}
\Eeqa
Here $c_{i,j}$ are the coupling constants.
Next we make
the Fourier transformation for multipolar operator of Np ions,
$O_{\gamma}(\BR_{\ell})
=\frac{1}{\sqrt{N}}\sum_{\rm q}O_{\gamma,\Bq}
{\rm e}^{{\rm i}\Bq\cdot\BR_{\ell}}
$,
where $\BR_{\ell}$ denotes the site of Np ion, and $\gamma$ denotes the symmetry
of the multipolar operator.

The interaction for nuclear spins at 
$(\frac{a}{2},\frac{a}{2},\frac{3a}{2})$,
$(\frac{3a}{2},\frac{a}{2},\frac{a}{2})$ and 
$(\frac{a}{2},\frac{3a}{2},\frac{a}{2})$  can be 
obtained by translating the site by 
$(-a,-a,0)$,
$(0,-a,-a)$ and $(-a,0,-a)$, respectively.
While the interaction of nuclear spins at
$(\frac{3a}{2},\frac{3a}{2},\frac{a}{2})$,
$(\frac{a}{2},\frac{a}{2},\frac{a}{2})$,
$(\frac{3a}{2},\frac{a}{2},\frac{3a}{2})$ and 
$(\frac{a}{2},\frac{3a}{2},\frac{3a}{2})$ can be 
obtained by 90 degree rotation around z axis of the T$_{\rm d}$ and
subsequent translation of the site by 
$(0,0,-a)$,
$(-a,-a,-a)$, $(0,-a,0)$ and $(-a,0,0)$, respectively.
By the rotation, the operators are changed as follows: 
$(I_{x},I_{y},I_{z}) \rightarrow (I_{y},-I_{x},I_{z})$,
$(J_{x},J_{y},J_{z}) \rightarrow (J_{y},-J_{x},J_{z})$,
$(T^{\beta}_{x},T^{\beta}_{y},T^{\beta}_{z})
 \rightarrow (-T^{\beta}_{y},T^{\beta}_{x},-T^{\beta}_{z})$
and $T_{xyz} \rightarrow -T_{xyz}$.
The sites are changed as
$(111) \rightarrow (1\bar{1}1)$,
$(1\bar{1}\bar{1}) \rightarrow (11\bar{1})$,
$(\bar{1}1\bar{1}) \rightarrow (\bar{1}\bar{1}\bar{1})$ and
$(\bar{1}\bar{1}1) \rightarrow (1\bar{1}1)$.

The Fourier transformation is carried out in the same way as was 
done for
eq. (\ref{eqA.2}) of the  $\brho_{1}$ site. 
Note that the same expression can be  obtained by 
transforming the wave vector in 
the Fourier transformed equation for $\brho_{1}$, i. e. 
$
(q_{x},q_{y},q_{z}) 
\rightarrow
(q_{y},-q_{x},q_{z}) 
$
except for  the factor,
$\EX^{\I\Bq\brho}$.
Of course, the operators are changed in the way noted above.

\section{Quadrupolar coupling in NpO$_{2}$
}

Let us consider the quadrupolar interaction between 
the $\Gamma_{3}$-type nuclear quadrupolar moment of O nucleus and the
$\Gamma_{5}$-type quadrupolar moment of Np ions,
\Beqa
H_{\rm qq}(\brho_{1},c_{2,2})
\nonumber \\
 & = &
c_{2,2}
\Big[
O_{u}(\brho_{1})
\frac{1}{\sqrt{4}\sqrt{6}}
\Big\{ ( O_{yz}+O_{zx}-2O_{xy})_{(111)}
 +( O_{yz}-O_{zx}+2O_{xy})_{(1\bar{1}\bar{1})}
\nonumber \\
 & &
 +(-O_{yz}+O_{zx}+2O_{xy})_{(\bar{1}1\bar{1})}
 +(-O_{yz}-O_{zx}-2O_{xy})_{(\bar{1}\bar{1}1)}
\Big\}
\nonumber \\
 & &
+O_{u}(\brho_{1})
\frac{1}{\sqrt{4}\sqrt{2}} 
\Big\{ (-O_{yz}+O_{zx})_{(111)}
     +(-O_{yz}-O_{zx})_{(1\bar{1}\bar{1})}
\nonumber \\
 & &
     +( O_{yz}+O_{zx})_{(\bar{1}1\bar{1})}
     +( O_{yz}-O_{zx})_{(\bar{1}\bar{1}1)}
\Big\}
\Big].
\Eeqa
Carrying out the Fourier transformation, we get
\Beqa
H_{\rm qq}(\brho_{1},c_{2,2})
\nonumber \\
 & = &
 \EX^{\I\Bq\brho_{1}}c_{2,2}\frac{8}{\sqrt{4}\sqrt{6}}
\Big[
\Big\{ \frac{1}{2}(
-O_{u}(\brho_{1})+\sqrt{3}O_{v}(\brho_{1}))O_{yz,\Bq}
(-\I s_{x}c_{y}c_{z}+c_{x}s_{y}s_{z})
\Big\}
\nonumber \\
 & &
+\Big\{ O_{u}(\brho)O_{xy,\Bq}
(-\I s_{z}c_{x}c_{y}+c_{z}s_{x}s_{y})
\Big\}
\nonumber \\
 & &
+
\Big\{ \frac{1}{2}(
-O_{u}(\brho_{1})-\sqrt{3}O_{v}(\brho_{1}))O_{zx,\Bq}
(-\I s_{y}c_{z}c_{x}+c_{y}s_{z}s_{x})
\Big\}
\Big].
\Eeqa
This corresponds to the interaction term with the coupling 
term $c_{2,2}$ in eq. (\ref{eq2.2}).

\section{Derivation of Invariant Coupling Form in CeB$_{6}$ 
}

Let us consider the invariant coupling form between the nuclear spin $\BI$ of B
on $(\frac{a}{2},\frac{a}{2},ua)$ and multipolar moments of f-electrons of Ce ions at
$(a,a,a)$, $(0,a,a)$, $(0,0,a)$ and $(a,0,a)$
of the cube of the s.c. lattice. 
For simplicity we denote these Ce ions as
$
(111), (\bar{1}11), (\bar{1}\bar{1}1),
$ 
and
$(1\bar{1}1)$, respectively.

These ions have the C$_{\rm 4v}$ symmetry around the z axis.
The nuclear spin operator $I_{z}$ of B belongs to the A$_{2}$ 
representation, and 
($I_{x},I_{y}$) form a two-dimensional E representation.

We can construct combinations of multipolar operators of the 
Ce ions, which belong to  A$_{2}$ and E representations.
Then, the invariant form of the hyperfine interaction of bilinear type is obtained as follows:
\Beqa
H_{\rm hf}(\frac{a}{2},\frac{a}{2},ua)
 & = &
I_{z}\Big[ 
c_{1,1}\frac{1}{\sqrt{4}}
\Big\{
 (J_{z})_{(111)}
+(J_{z})_{(\bar{1}11)}
+(J_{z})_{(\bar{1}\bar{1}1)}
+(J_{z})_{(1\bar{1}1)}
\Big\}
\nonumber \\
& & 
+c_{1,2}\frac{1}{\sqrt{8}}
\Big\{
 ( J_{x}+J_{y})_{(111)}
+(-J_{x}+J_{y})_{(\bar{1}11)}
+(-J_{x}-J_{y})_{(\bar{1}\bar{1}1)}
+( J_{x}-J_{y})_{(1\bar{1}1)}
\Big\}
\nonumber \\
& &
+c_{1,3}\frac{1}{\sqrt{8}}
\Big\{
 ( T^{\beta}_{x}-T^{\beta}_{y})_{(111)}
+(-T^{\beta}_{x}-T^{\beta}_{y})_{(\bar{1}11)}
+(-T^{\beta}_{x}+T^{\beta}_{y})_{(\bar{1}\bar{1}1)}
+( T^{\beta}_{x}+T^{\beta}_{y})_{(1\bar{1}1)}
\Big\}
\nonumber \\
& &
+c_{1,4}\frac{1}{\sqrt{4}}
\Big\{
 ( T_{xyz})_{(111)}
+(-T_{xyz})_{(\bar{1}11)}
+( T_{xyz})_{(\bar{1}\bar{1}1)}
+(-T_{xyz})_{(1\bar{1}1)}
\Big\}
\Big]
\nonumber \\
& + &
I_{x}
\Big[
c_{2,1}\frac{1}{\sqrt{4}}
\Big\{
 ( J_{x})_{(111)}
+(J_{x})_{(\bar{1}11)}
+(J_{x})_{(\bar{1}\bar{1}1)}
+( J_{x})_{(1\bar{1}1)}
\Big\}
\nonumber \\
& &
+c_{2,2}\frac{1}{\sqrt{4}}
\Big\{
 ( J_{z})_{(111)}
+(-J_{z})_{(\bar{1}11)}
+(-J_{z})_{(\bar{1}\bar{1}1)}
+( J_{z})_{(1\bar{1}1)}
\Big\}
\nonumber \\
& &
+c_{2,3}\frac{1}{\sqrt{4}}
\Big\{
 ( J_{y})_{(111)}
+(-J_{y})_{(\bar{1}11)}
+( J_{y})_{(\bar{1}\bar{1}1)}
+(-J_{y})_{(1\bar{1}1)}
\Big\}
\nonumber \\
& &
+c_{2,4}\frac{1}{\sqrt{4}}
\Big\{
 ( T^{\beta}_{x})_{(111)}
+( T^{\beta}_{x})_{(\bar{1}11)}
+( T^{\beta}_{x})_{(\bar{1}\bar{1}1)}
+( T^{\beta}_{x})_{(1\bar{1}1)}
\Big\}
\nonumber \\
& &
+c_{2,5}\frac{1}{\sqrt{4}}
\Big\{
 (-T^{\beta}_{y})_{(111)}
+( T^{\beta}_{y})_{(\bar{1}11)}
+(-T^{\beta}_{y})_{(\bar{1}\bar{1}1)}
+( T^{\beta}_{y})_{(1\bar{1}1)}
\Big\}
\nonumber \\
& &
+c_{2,6}\frac{1}{\sqrt{4}}
\Big\{
 ( T^{\beta}_{z})_{(111)}
+(-T^{\beta}_{z})_{(\bar{1}11)}
+(-T^{\beta}_{z})_{(\bar{1}\bar{1}1)}
+( T^{\beta}_{z})_{(1\bar{1}1)}
\Big\}
\nonumber \\
& &
+c_{2,7}\frac{1}{\sqrt{4}}
\Big\{
 ( T_{xyz})_{(111)}
+( T_{xyz})_{(\bar{1}11)}
+(-T_{xyz})_{(\bar{1}\bar{1}1)}
+(-T_{xyz})_{(1\bar{1}1)}
\Big\}
\Big]
\nonumber \\
&+ &
 I_{y}
\Big[
c_{2,1}\frac{1}{\sqrt{4}}
\Big\{
 (J_{y})_{(111)}
+(J_{y})_{(\bar{1}11)}
+(J_{y})_{(\bar{1}\bar{1}1)}
+(J_{y})_{(1\bar{1}1)}
\Big\}
\nonumber \\
& &
+c_{2,2}\frac{1}{\sqrt{4}}
\Big\{
 ( J_{z})_{(111)}
+( J_{z})_{(\bar{1}11)}
+(-J_{z})_{(\bar{1}\bar{1}1)}
+(-J_{z})_{(1\bar{1}1)}
\Big\}
\nonumber \\
& &
+c_{2,3}\frac{1}{\sqrt{4}}
\Big\{
 ( J_{x})_{(111)}
+(-J_{x})_{(\bar{1}11)}
+( J_{x})_{(\bar{1}\bar{1}1)}
+(-J_{x})_{(1\bar{1}1)}
\Big\}
\nonumber \\
& &
+c_{2,4}\frac{1}{\sqrt{4}}
\Big\{
 ( T^{\beta}_{y})_{(111)}
+( T^{\beta}_{y})_{(\bar{1}11)}
+( T^{\beta}_{y})_{(\bar{1}\bar{1}1)}
+( T^{\beta}_{y})_{(1\bar{1}1)}
\Big\}
\nonumber \\
& &
+c_{2,5}\frac{1}{\sqrt{4}}
\Big\{
 ( T^{\beta}_{x})_{(111)}
+(-T^{\beta}_{x})_{(\bar{1}11)}
+( T^{\beta}_{x})_{(\bar{1}\bar{1}1)}
+(-T^{\beta}_{x})_{(1\bar{1}1)}
\Big\}
\nonumber \\
& &
+c_{2,6}\frac{1}{\sqrt{4}}
\Big\{
 (-T^{\beta}_{z})_{(111)}
+(-T^{\beta}_{z})_{(\bar{1}11)}
+( T^{\beta}_{z})_{(\bar{1}\bar{1}1)}
+( T^{\beta}_{z})_{(1\bar{1}1)}
\Big\}
\nonumber \\
& &
+c_{2,7}\frac{1}{\sqrt{4}}
\Big\{
 ( T_{xyz})_{(111)}
+(-T_{xyz})_{(\bar{1}11)}
+(-T_{xyz})_{(\bar{1}\bar{1}1)}
+( T_{xyz})_{(1\bar{1}1)}
\Big\}
\Big]
\ .
\label{eqA.1}
\Eeqa
Here $c_{i,j}$ are coupling constants.
By carrying out the Fourier transformation,
we get the interaction form given in eq.({\ref{eq4.1}).

The interaction for the nuclear spin of B on 
$(\frac{a}{2},\frac{a}{2},-ua)$ can be obtained 
by a mirror operation which moves B on $(\frac{a}{2},\frac{a}{2},ua)$
to 
 $(\frac{a}{2},\frac{a}{2},-ua)$.  
By this transformation the multipolar operators are changed as follows:
$
(I_{x},I_{y},I_{z}) \rightarrow (-I_{x},-I_{y},I_{z}), \H
(J_{x},J_{y},J_{z}) \rightarrow (-J_{x},-J_{y},J_{z}), \H
(T^{\beta}_{x},T^{\beta}_{y},T^{\beta}_{z})
\rightarrow 
(-T^{\beta}_{x},-T^{\beta}_{y},T^{\beta}_{z})
$
and 
$
T_{xyz} \rightarrow T_{xyz}
$.
At the same time, the sites are changed as
$
(111) \rightarrow (11\bar{1}), \H
(\bar{1}11) \rightarrow (\bar{1}1\bar{1}),\H
(\bar{1}\bar{1}1) \rightarrow (\bar{1}\bar{1}\bar{1}),\H
$
and 
$
(1\bar{1}1) \rightarrow (1\bar{1}\bar{1})
$.
The Fourier transformation is carried out in the same way as was done to derive
eq. (\ref{eq4.1}). Note here that the same expression can be  obtained by 
transforming the wave vector in eq. (\ref{eq4.1}) as
$
(q_{x},q_{y},q_{z}) 
\rightarrow
(q_{x},q_{y},-q_{z}) 
$
except for  the factor,
$\EX^{\frac{q_{x}+q_{y}+q_{z}}{2}}\EX^{\I\Bq\brho}$. 

\vfill\eject


\begin{thebibliography}{9}
\bibitem{E1}
In the strict limit of the zero field, the level structure 
of the  quadrupolar splitting 
of non-cubic sites will be equivalent though the principal axes are 
different with each other.
Therefore there is the possibility that the quadrupolar splitting can be 
observed even in the polycrystalline samples.
This is pointed out by S. Kambe.
\end{thebibliography}
\end{document}